# Multipath Interference Suppression of Amplitude-Modulated Continuous Wave Coaxial-Scanning LiDAR Utilizing Bayesian-Optimized XGBoost Ensemble

Sung-Hyun Lee, Yoon-Seop Lim, Wook-Hyeon Kwon, Yong-Hwa Park

*Abstract*— This paper proposes a novel multipath interference (MPI) suppression method in the amplitude-modulated continuous wave (AMCW) coaxial scanning LiDAR. Previous works have focused on the MPI suppression in conventional AMCW time-of-flight (ToF) sensors with flash type illumination sources based on the various MPI assumptions, whose MPI errors remain cm-scale. To achieve mm-scale MPI error suppression, this paper proposes a novel MPI error suppression method implemented in a coaxial type AMCW scanning LiDAR in which the MPI phenomenon can be accurately modelled. The proposed method utilizes this novel MPI mathematical model in conjunction with Bayesian-optimized extreme gradient boosting (XGBoost) ensemble. MPI synthetic dataset generated by the proposed MPI model is used in the training of the XGBoost ensemble. Experimental validation showed that the mean absolute error (MAE) of MPI error can be reduced to less than 2 mm by the proposed method. Such precise MPI suppression results are also maintained in real object scenes. Specifically, the MAE of MPI error in object scene with sharp corner is reduced to 2.8 mm, which is extremely low compared to other previous works.

## I. INTRODUCTION

DETPH information is widely used nowadays for the precise object recognition of intelligent mechatronic systems. Autonomous robots and vehicles generally utilize 3D spatial information extracted from depth map for their localization, indoor mapping, and obstacle detection [1], [2]. Such 3D depth information is also utilized to provide human pose information for many engineering applications [3]. Likewise, the utilization of 3D depth information is already a main trend for the visual recognition of the state-of-the art intelligent systems and devices.

Amplitude-modulated continuous wave (AMCW) time-of-flight (ToF) sensor is one of the typical 3D depth measurement devices [4]–[8]. The main principle of the AMCW ToF sensor is to measure the phase delay of modulated laser signal reflected from an object using demodulation pixel. The majority of AMCW ToF sensors adopt flash type illumination optics with CMOS demodulation pixel arrays [4], [6], [7].

However, there exist some problems related to systematic distance measurement errors such as $2\pi$-ambiguity [9], fixed pattern noise [4], [10], thermal fluctuation [4], [10], etc. These systematic errors can be normally mitigated by utilizing look-up table (LUT) and various spatial image processing methods, according to many previous research works [4], [10]. On the other hand, multipath interference (MPI), which is a non-systematic error, is still a challenging problem [11]–[18].

Since the illuminated light is spread over the entire object scene by diffuser, multiple light rays from unwanted region can also enter demodulation pixels. Since it is not feasible to get rid of the MPI directly, many researchers have attempted to reduce the MPI error using various post-error correction methods. One of the MPI suppression methods is based on the optimization scheme in conjunction with various MPI models. Dorrington *et al.* [11] assumed two-path model for MPI, and designed an objective function with $L_2$ norm to find true depth using least-square approach. Kirmani *et al.* [12] assumed multipath model as $N$ discrete paths, and estimated multipath parameters using the total least square (TLS) Prony's method. Meanwhile, Freedman *et al.* [13] used sparse reflections analysis (SRA) method to validate the two-path MPI model with data measured in the condition of 3 different modulation frequencies. With similar concepts, Feigin *et al.* [14] used matrix pencil method which is one of the spectral estimation methods to find out true depth from the data measured with 3 different modulation frequencies. All these aforementioned previous works assume MPI models as two or more number of paths of light rays, and find the true depth based on the convex optimization or spectral estimation. However, the optimization solver is relatively slow making real-time implementation hard. Moreover, although two-path MPI model quite fits well with many conventional AMCW ToF sensors, there still exist lots of redundant rays of light in actual MPI situation.

To confront such imperfect light transport model and slow computation process, many researchers have utilized post-correction models based on deep learning approaches recently. The majority of deep learning-based approaches generally adopt the convolutional neural network (CNN) architectures [15]–[18]. Marco *et al.* [15] proposed a convolutional autoencoder (CAE) architecture to train a synthetic MPI data set. Agresti *et al.* [16] suggested a coarse-to-fine CNN architecture to train the synthetic MPI data set generated by a generative adversarial networks (GAN) with small amount of

Sung-Hyun Lee, Yoon-Seop Lim, and Yong-Hwa Park are with the Department of Mechanical Engineering, Korea Advanced Institute of Science and Technology, Republic of Korea. (*Corresponding author*: Yong-Hwa Park.)

Wook-Hyeon Kwon is with the Mechatronics Research, Samsung Electronics Co., Republic of Korea.

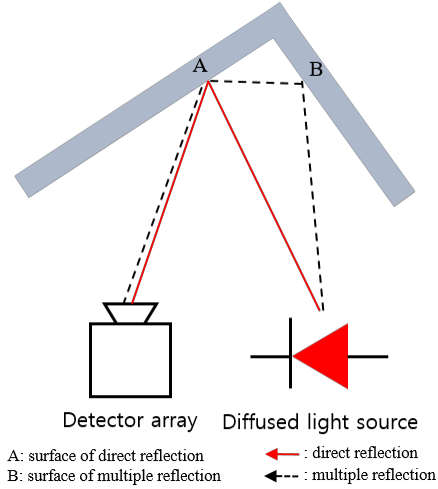

Fig. 1. Multipath interference in conventional AMCW ToF sensor. The red line is directly reflected light. The dotted black line is multi-reflected light. Since the light source is flash-type, the B is widely distributed over the entire counter facet of A.

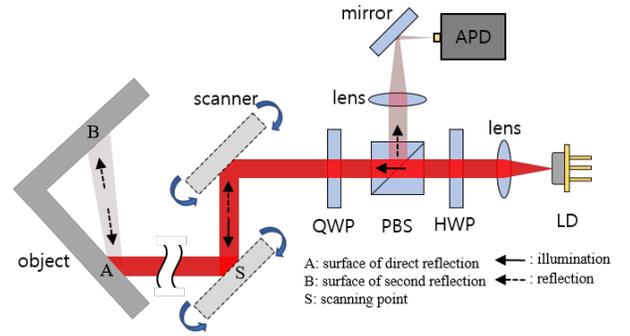

Fig. 2. Multipath interference in coaxial type AMCW scanning LiDAR proposed in this paper. QWP is quarter-wave plate. PBS is polarizing beam splitter. HWP is half-wave plate. LD is laser diode. APD is avalanche photodiode. Emitted light from LD and reflected light to APD have common (coaxial) path after PBS. Since the laser beam is Gaussian beam and collimated almost point-like in A, the majority of reflected light rays at B is also concentrated on relatively narrow region compared to Fig. 1.

real MPI dataset. Unlike other CNN-based research works, Su *et al.* [17] used intermediate raw amplitude image for training the CNN architecture, not depth image. As mentioned above, many research works for MPI suppression mainly use 3D depth and amplitude images with multiple modulation frequencies to train CNN-based deep learning architecture. Some research works use simple fully connected layer network (FCN) without convolution layer to train MPI data, but such cases are rare [18]. The aforementioned CNN-based approaches can reduce the MPI error to cm-scale by training the geometrical pattern of measured scene with MPI. However, generating accurate synthetic data set is still a challenging issue since the amount of labelled real MPI images is extremely limited. Moreover, as CNN architecture entirely depends on the geometrical patterns, the MPI reduction performance is not robust against the orientation and distance of measured scene even for the same object.

In this paper, regarding the above mentioned issues, a novel MPI suppression method in both hardware and software levels is proposed and experimentally validated. To reduce the inherent MPI error in hardware level, coaxial type scanning optics is combined with AMCW scheme [8], [19], [20]. Since the optical path of received MPI light should share the path of emitted light from LiDAR, the relative ratio of MPI light in total received light of photodetector is much less than that of the conventional ToF sensors which utilize flash-type illumination sources shown as Fig. 1. Consequently, the coaxial scanning optics can maintain high optical SNR, in other words, relatively low MPI error compared to the conventional flash-type ToF sensors. To further reduce such MPI error in software level, a pixel-wise nonlinear mapping function which matches raw input depth and amplitude with scalar output of true depth is designed based on Bayesian-optimized eXtreme Gradient Boosting (XGBoost) ensemble [21], [22]. Unlike previous CNN-based architectures trained with image inputs, a novel pixel-wise learning approach is newly adopted to avoid the dependency on the geometrical patterns of measured scenes with MPI. Meanwhile, to tackle the issue related to the generation of training data, an exact physical model of the MPI in the AMCW coaxial scanning LiDAR is also derived and utilized to generate enough synthetic MPI dataset. The exact synthetic MPI dataset with 4 different modulation frequencies are then utilized in training of the proposed XGBooost ensemble with optimized hyperparameters [23], [24]. For the hyperparameter optimization, tree-structured Parzen estimator (TPE) in Bayesian scheme is used. The proposed XGBoost-based MPI suppression algorithm can reduce the MPI error to mm-scale according to the validation results in this paper. Some previous research works related to the AMCW scanning LiDAR had ignored such MPI effect [25], [26]. To our best knowledge, this paper is the first case study for the MPI suppression in coaxial type AMCW scanning LiDAR. We anticipate that the proposed pixel-wise learning approach with precise synthetic data can be one of solutions to reduce the MPI error of various coaxial type scanning LiDAR, not just confined to AMCW scheme.

## II. Problem Statement of Multipath Interference in AMCW ToF Sensors

Fundamental principle of AMCW ToF measurement is to estimate the time delay (a.k.a. phase delay) of reflected light signal relative to a reference signal, *i.e.*, demodulation signal. These signals can be expressed as follows [4], [6], [8]:

$$r(t) = R \cdot \left(1 + \alpha \cdot \cos\left(w\left(t - \frac{2d}{c}\right)\right)\right) \quad (1)$$

$$s(t) = m \cdot \cos(wt) \quad (2)$$

$$w = 2\pi f \quad (3)$$

where $r(t)$ is the reflected light signal, $\alpha$ is modulation contrast, $R$ is the reflectivity of scene, $d$ is the distance of the scene from the sensor, $c$ is the velocity of light, $s(t)$ is the demodulation signal, $m$ is the amplitude of $s(t)$, and $f$ is the modulation frequency. To estimate the time (phase) delay, the reflected light signal is demodulated with at least three different time (phase)-shifted demodulation signals. This process can be expressed functionally using cross correlation as follows:

$$C(\tau_n) = \frac{1}{T_{int}} \int_0^{T_{int}} r(t) \cdot s(t+\tau_n) dt$$
$$= \frac{1}{T_{int}} \int_0^{T_{int}} R \cdot \left(1 + \alpha \cdot \cos\left(w\left(t - 2d/c\right)\right)\right)$$
$$\times m \cdot \cos\left(w(t+\tau_n)\right) dt$$
$$= R \cdot \frac{\alpha \cdot m}{2} \cos\left(w\tau_n + \phi(w)\right) (n=1,2,...N_{tap}) \quad (4)$$

$$\phi(w) = \arctan\left(\frac{C(\tau_3) - C(\tau_1)}{C(\tau_0) - C(\tau_2)}\right) \quad (5)$$

$$\Gamma = \frac{\sqrt{(C(\tau_3) - C(\tau_1))^2 + (C(\tau_2) - C(\tau_0))^2}}{2} \quad (6)$$

$$d = \frac{c \cdot \phi(w)}{4\pi \cdot f} \quad (7)$$

where $C(\tau_n)$ is the $n$-th cross correlation, $\tau_n$ is the time shift of the $n$-th demodulation signal, $N_{tap}$ is the number of taps which is 4 in most cases, $T_{int}$ is the integration time, $\Gamma$ is the amplitude of cross correlation, and $\phi(w)$ is the phase delay. By sampling 4 cross correlations with 4 time shifts of demodulation signals in general, the phase delay $\phi(w)$ can be identified as (5) [6]–[8]. Meanwhile, the phasor expression can be effectively used in order to express the cross correlation with only its amplitude and phase delay as follow [13], [14]:

$$\mathbf{C}_n(w) = R \cdot \frac{\alpha \cdot m}{2} e^{jw\tau_n} e^{j\phi(w)} = \mathbf{\Gamma}_n \cdot e^{j\phi(w)} \quad (8)$$

where $\mathbf{\Gamma}_n$ is the complex-valued amplitude of cross correlation, $\phi(w)$ is the phase delay of the cross correlation, and $\mathbf{C}_n(w)$ is the phasor expression of (4). Without loss of generality, the phasor expression in (8) can be simplified by considering only 0-phase shift as a representative phasor expression as follow [13], [14], [27]:

$$\mathbf{C}(w) = \mathbf{C}_0(w) = R \cdot \frac{\alpha \cdot m}{2} e^{j\phi(w)} = \Gamma \cdot e^{j\phi(w)} \quad (9)$$

where $\Gamma$ is the amplitude of cross correlation defined as (6). All cross correlations in following derivations are expressed using (9) in this paper.

In ideal case, per demodulation pixel, only directly reflected light signal from a specific object point is matched. However, due to the multiple reflection(s) of the illuminated light, other unwanted light signals also enter the demodulation pixel as shown in Fig. 1 in case of flash-type ToF sensors. Consequently, the measured cross correlation function at a demodulation pixel is the summation of multiple complex phasors as follow [14], [27]:

$$C_{net}(w) = \Gamma_D \cdot e^{j\phi_D(w)} + \sum_{k=1}^{M-1} \Gamma_k \cdot e^{j\phi_k(w)} \quad (10)$$

where $\Gamma_D$, $\phi_D$ are the amplitude and phase delay of cross correlation with directly reflected light, $\Gamma_k$, $\phi_k$ are the amplitude and phase delay of cross correlation with the $k$-th multi-reflected light, and $M$ is the total number of light paths including direct reflection. As shown in (10), the measured cross correlation inevitably includes multiple unknown complex phasors of which modulation frequencies are all same. From the trigonometric property, the summation of all complex phasors in (10) is resultantly same as another single complex phasor with a distorted phase delay. Consequently, it is not feasible to obtain $\phi_D$ using only single modulation frequency due to the redundancy of unknown parameters in (10). To correctly estimate $\phi_D$ in (10), measured cross correlation samples with multiple modulation frequencies are needed. The required number of modulation frequencies is determined by the MPI model, *i.e.*, the total number of amplitudes and phase delays in (10). Dorrington *et al.* [11] used two-path assumption ($M=2$) in which the number of unknown parameters is 4. Since the intensity of MPI actually decays exponentially as the length of MPI path increases, it is quite reasonable to assume the MPI model using 2 paths of light [12], [13], [27]. Other previous research works also assume the number of MPI path as 2 or 3 in general [11], [13], [14], [27]–[29]. Based on such assumed MPI models, many research works had improved the performance of MPI error reduction using various optimization methods and deep learning architectures.

However, there still exist limitations in previous works dealing with MPI. First, many researchers used less than 3 modulation frequencies although at least 4 modulation frequencies are required for two-path model. This is mainly due to the hardware limitation of conventional ToF sensors. Second, the complexity of actual MPI in conventional AMCW ToF sensor is much higher than simple two-path model. Namely, achieving precise model of the MPI in conventional ToF sensor is quite cumbersome due to the optical characteristics of flash type illumination source and pixel array resulting in low SNR of directly reflected light. At last, for the learning-based approaches, the lack of synthetic dataset also affects the MPI suppression results in negative, causing over fitting. Due to the aforementioned limitations, the MPI suppression results of previous works still remain cm-scale.

To cope with aforementioned limitations, a pixel-wise learning approach based on Bayesian-optimized XGBoost is proposed and implemented in AMCW scanning LiDAR to reduce MPI error in this paper [8], [21], [22]. Additionally, to increase the information of directly reflected light, 4 different modulation frequencies are adopted. Meanwhile, to train the XGBoost-based pixel-wise MPI correction algorithm with enough and precise data, simulation model with the sensor noise characteristics of AMCW scanning LiDAR is proposed. Due to the simplicity of coaxial optics and single avalanche photodiode (APD), simulation modeling including MPI is much easier compared to conventional ToF sensors [8]. By training the XGBoost ensemble with precise MPI simulation data per pixel, the MPI error is reduced to mm-scale in various object scenes according to the validation results in this paper.

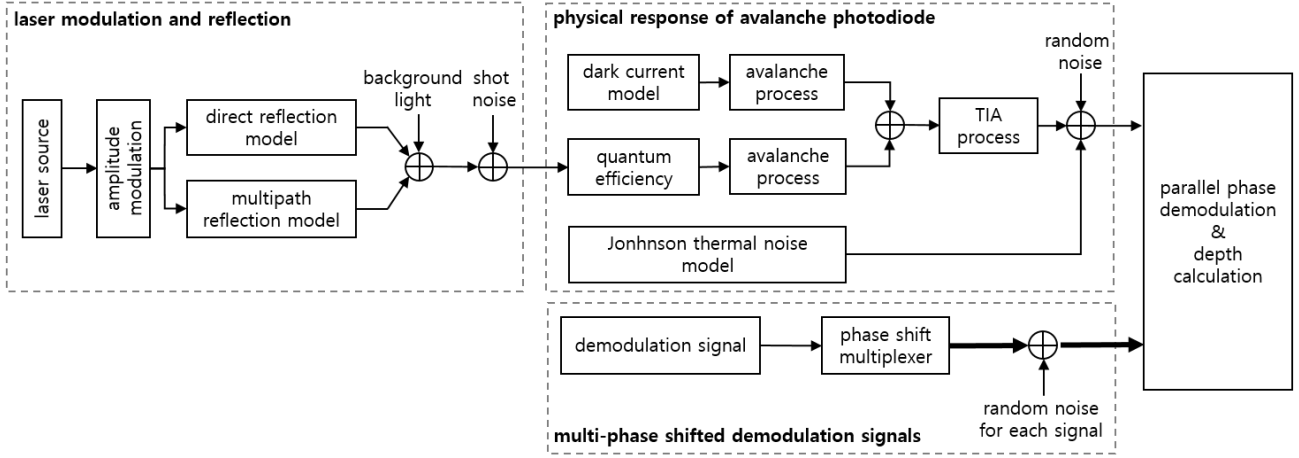

Fig. 3. Block diagram of sensor simulation model for AMCW coaxial scanning LiDAR based on parallel-phase demodulation [8].

### III. PIXEL-WISE MPI ERROR LEARNING APPROACH USING PRECISE MPI MODEL AND XGBOOST ENSEMBLE

#### A. Light Transport Model in AMCW Coaxial Scanning LiDAR

The coaxial type scanning LiDAR in Fig. 2 has mainly two different aspects compared to the biaxial flash type ToF sensors in Fig. 1. First, since the laser beam is collimated in very narrow point A (mm-scale) as shown in Fig. 2, the surface area of B on which multi-reflected light rays are mainly distributed can be roughly considered relatively narrow compared to Fig. 1. Additionally, as the power distribution profile of the used laser in Fig. 2 is Gaussian beam, most multi-reflected light rays are mainly distributed on the center of B. Second, the feasible MPI optical path is only confined to the case of triple reflection in the sequence of A, B, and A which also shares the emitted path from S to A in Fig. 2. Considering aforementioned characteristics, the two-path model assumption which were widely used in many previous works related to flash-type ToF sensors [11], [13], [14], [27]–[29] can be much more reasonably adopted in Fig. 2. Namely, the total measured cross correlation in (10) can be summarized as the summation of each cross correlation corresponding to directly reflected light from A to S and MPI light which additionally has round trip between center of A and center of B, as follow:

$$C_{net}(w) = \Gamma_D \cdot e^{j\phi_D(w)} + \Gamma_M \cdot e^{j\phi_M(w)} \quad (11)$$

where $\Gamma_D$, $\phi_D$ are the amplitude and phase delay of cross correlation with directly reflected light, $\Gamma_M$, $\phi_M$ are the amplitude and phase delay of cross correlation with multi-reflected light in Fig. 2. The amplitude and phase delay of each cross correlation in (11) can be modeled based on the radiometric properties [30], [31] as follows:

$$\Gamma_D = \Gamma_R \cdot \frac{\rho_{SAS}}{\|d_{AS}\|^2} \quad (12)$$

$$\Gamma_M = \Gamma_R \cdot \frac{(\rho_{SAB} \cdot \rho_{ABA} \cdot \rho_{BAS})}{\|d_{AS}\|^2} \quad (13)$$

$$\phi_D(w) = \frac{4\pi \cdot f \cdot d_{AS}}{c} \quad (14)$$

$$\phi_M(w) = \frac{4\pi \cdot f \cdot (d_{AS} + d_{AB})}{c} \quad (15)$$

where $\Gamma_R$ is the initial value of cross correlation, $d_{XY}$ is the absolute distance between center of X and center of Y, $\rho_{XYZ}$ is the bidirectional reflectance distribution function (BRDF) of which the direction is from X to Z through Y. Aforementioned amplitude equations are based on the fact that the received light intensity is proportional to the reflectivity and inversely proportional to the square of travel length [30], [31]. By controlling the model parameters in (12) to (15), numerous received light signals including MPI in simulation level can be synthesized and utilized to make precise synthetic MPI dataset reflecting the physics of light transport. Meanwhile, such light transport model in (11) is combined with sensor response model including electronic noise characteristics for precise simulation of AMCW coaxial scanning LiDAR measurement in following subsection.

#### B. Sensor Simulation model of AMCW Coaxial Scanning LiDAR Based on Parallel-Phase Demodulation Including Noise Characteristics

The total received light signal including directly reflected light and MPI is generated with the light transport model in previous section. After this light signal is sensed by APD, various physical processes occur inside the photoelectric circuits of APD. Such sensor response should be considered for the precise AMCW coaxial scanning LiDAR simulation model. To reflect such physical phenomena, sensor noise characteristics are considered in the sensor simulation model as shown in Fig. 3.

In Fig. 3, after the laser source is amplitude-modulated in sinusoidal waveform as (1), the laser signal is reflected following the light transport model in (11). Then the number of received photons in incident laser signal is expressed as follows [32]:

$$n_{ph} = round\left(\frac{P_{opt} \cdot t_{transit}}{E_p}\right) \quad (16)$$

$$E_p = \frac{hc}{\lambda} \quad (17)$$

where $n_{ph}$ is the number of received photons, $P_{opt}$ is the power of incident received light, $t_{transit}$ is the carrier transit time, $E_p$ is the energy of a photon, $h$ is the Planck constant, and $\lambda$ is wavelength of laser. However, the actual number of received photons is determined in stochastic way due to the photon shot noise as follow [32]:

$$n_{ph,shot} \sim P_{Poisson}(n_{ph,shot};n_{ph}) = \frac{n_{ph}^{n_{ph,shot}} e^{-n_{ph}}}{n_{ph,shot}!} \quad (18)$$

where $n_{ph,shot}$ is the number of received photons with shot noise, and $P_{Poisson}()$ is probability density function (PDF) with Poisson distribution. The received photons are then converted into electrons in APD and amplified by avalanche process as follows [32]:

$$n_{e^-} = n_{ph,shot} \cdot \eta + \varepsilon_{back} \quad (19)$$

$$\overline{n}_{ph,apd} = M \cdot n_{e^-} \quad (20)$$

$$\mathrm{var}(n_{ph,apd}) = M^2 F \cdot n_{e^-} \quad (21)$$

$$n_{ph,apd} \sim N_{avalanche}\left(n_{ph,apd};\overline{n}_{ph,apd},\mathrm{var}(n_{ph,apd})\right) \quad (22)$$

where $n_{e^-}$ is the number of photoelectrons, $\eta$ is quantum efficiency, $\varepsilon_{back}$ is additional Gaussian noise due to background light, $M$ is avalanche gain, $F$ is excess noise factor, $n_{ph,apd}$ is amplified number of electrons after avalanche process, $\overline{n}_{ph,apd}$ is the average of $n_{ph,apd}$, $\mathrm{var}(n_{ph,apd})$ is the variance of $n_{ph,apd}$, and $N_{avalanche}()$ is the PDF of avalanche process which is same as Gaussian distribution. Such photoelectrons after avalanche process directly results in the photocurrent and corresponding voltage in APD as follows [33]–[35]:

$$I_{ph,apd} = \frac{n_{ph,apd} \cdot q}{t_{transit}} \quad (23)$$

$$V_{ph,apd} = I_{ph,apd} \cdot G \quad (24)$$

where $q$ is the unit charge of an electron, $I_{ph,apd}$ is the photocurrent generated by photoelectrons, $V_{ph,apd}$ is the corresponding voltage after amplification by transimpedance amplifier (TIA), and $G$ is the TIA gain.

Meanwhile, there exists another type of electric signal, *i.e.*, dark current, which is the electric noise mainly affected by the temperature of APD. The number of dark electrons with shot noise is expressed as follows [32], [36]:

$$n_{dark} = round\left(P_A I_{FM} T^{3/2} t_{transit} \exp\left(-\frac{E_g}{2k_B T}\right)\right) \quad (25)$$

$$n_{dark,shot} \sim P_{Poisson}(n_{dark,shot};n_{dark}) \quad (26)$$

where $n_{dark}$ is the number of dark electrons, $P_A$ is the active area of APD, $I_{FM}$ is dark current figure-of-merit, $T$ is temperature, $E_g$ is bandgap energy, $k_B$ is Boltzmann constant, and $n_{dark,shot}$ is the number of dark electrons with shot noise. The dark current in APD is then generated following (23) that replaces $n_{ph,apd}$ with $n_{dark,shot}$.

Except for the dark current, TIA noise is generated during TIA process [33]. Additionally, thermal noise is also generated due to the load of circuits. Such TIA noise and thermal noise are not negligible for the precise sensor response model. All these physical processes are described as follows [33], [37], [38]:

$$n_{TIA} \sim N_{TIA}\left(n_{TIA};\overline{n}_{TIA},\mathrm{var}(n_{TIA})\right) \quad (27)$$

$$\mathrm{var}(n_{TIA}) = \frac{t^2_{transit} S_{TIA} BW}{q^2} \quad (28)$$

$$I_{thermal} = \sqrt{\frac{4k_B T \cdot BW}{R_{Load}}} \quad (29)$$

where $n_{TIA}$ is the number of TIA noise electrons, $\overline{n}_{TIA}$ is the average of $n_{TIA}$, $\mathrm{var}(n_{TIA})$ is the variance of $n_{TIA}$, $N_{TIA}()$ is the PDF of $n_{TIA}$ same as Gaussian distribution, $S_{TIA}$ is the spectrum intensity of TIA noise, $BW$ is bandwidth, $I_{thermal}$ is the thermal noise current, and $R_{Load}$ is the load of circuit. In addition to thermal noise current in (29), the TIA noise current is also generated following (23) that replaces $n_{ph,apd}$ with $n_{TIA}$.

After aforementioned processes, each current results in corresponding voltage which is the product of current and TIA gain $G$. Consequently, the total voltage response of APD due to light signals and all noise signals can be presented as follows:

$$V_{apd} = V_{ph,apd} + V_{dark,apd} + V_{TIA} + V_{thermal} + \varepsilon_{rand} \quad (30)$$

where $V_{apd}$ is the total voltage response of APD, $V_x$ is the voltage generated by source $x$, and $\varepsilon_{rand}$ is additional random noise which is modeled as a simple Gaussian noise.

To demodulate (30), demodulation signal in (2) is generated in the simulation as shown in Fig. 3. By multiplexing the demodulation signal with multiple phase shifts (4 different phase shifts in most cases), the cross correlations are calculated in parallel in the sensor simulation [8]. Using these cross correlations, the synthetic depth and total amplitude of cross correlations are finally generated. By changing physical parameters of light transport model and APD response model, numerous synthetic MPI datasets can be generated to be used as training and validation dataset.

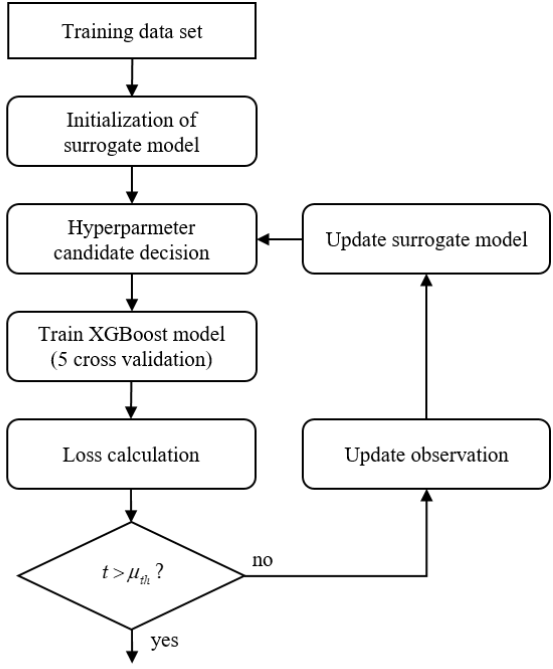

Fig. 4. Flow chart of the hyperparameter optimization based on TPE approach for XGBoost model.

*C. XGBoost Ensemble Optimized by TPE in Bayesian Scheme.*

XGBoost was developed in 2016 by Chen *et al.* [21]. The basic principle of XGBoost is same as the gradient boosting method which learns residuals using multiple weak learners. Based on this ground, the XGBoost was improved in the aspects of overfitting, training speed, and precision. Compared to the conventional boosting methods, the XGBoost prevents overfitting using regularization term in loss function. Meanwhile, to speed up the training process, the XGBoost trains each classification and regression tree (CART) in parallel. Additionally, by adding second-order Taylor expansion to the loss function, the mathematical precision was improved. Except for these logical improvements, XGBoost utilizes prefetching considering cache for improved data interface which can decrease the training time dramatically. Due to such advantages of XGBoost, many researchers have adopted the XGBoost to solve various data-driven engineering problems recently [23], [39]. In this paper, the XGBoost is adopted as MPI data training regressor to precisely correct MPI error in the AMCW coaxial scanning LiDAR. In the following paragraphs, the basic concepts and mathematical expressions of XGBoost are described.

The basic regression problem with $K$-CARTs are as follows:

$$\hat{y}_i = \sum_{t=1}^{K} f_t(x_i), f_t \in F \quad (31)$$

$$F = \left\{ f(x) = w_{q(x)} \mid w \in R^T \right\} \quad (32)$$

$$q : R^d \to \{1, 2, ..., T\} \quad (33)$$

$$D = \left\{ (x_i, y_i) \mid x \in R^d, y \in R \right\} (i = 1, 2, ..., n) \quad (34)$$

where $\hat{y}$ is estimated output by regressor, $f$ is weak learner which belongs to the category $F$, $K$ is the total number of CART learners, $F$ is the hypothesis space for all feasible CART learners, $T$ is the number of leaf nodes in specific CART, $d$ is the dimension of feature space, $w$ is the score vector of all leaf nodes, $q(x)$ is the mapping function which matches the input sample with leaf nodes of CART, $D$ is the training dataset, and $n$ is the total number of training data samples. To find optimal $K$-CARTs, the objective function is designed in XGBoost algorithm as follows:

$$J = \sum_{i=1}^{n} L_2(y_i, \hat{y}_i) + \sum_{t=1}^{K} \Omega(f_t) \quad (35)$$

$$\Omega(f_t) = \gamma T + \frac{1}{2} \lambda \sum_{j=1}^{T} w_j^2 \quad (36)$$

where $J$ is the total objective function of XGBoost, $L_2$ is the L2-norm loss composed of target value and estimated value, $\Omega$ is the regularization term of CART, $\gamma$ is the penalty coefficient related to the number of leaf node, and $\lambda$ is the penalty coefficient related to the square of weight. The training process of XGBoost is ultimately to minimize $J$. During training process, overfitting is prevented by the regularization term in (36). After minimization of (35), the optimal score vectors corresponding to each CART ensemble are finally obtained to be used for the XGBoost regression in (31) [21]. To estimate the correct distance, the measured depth and amplitude of cross correlation with 4 different modulation frequencies are utilized as input vector.

For the precise training of the XGBoost regressor, selection of proper hyperparameters is important. However, there exist many hyperparameters as shown above equations, which makes manual tuning difficult. To select optimal hyperparameters of XGBoost model, TPE-based Bayesian optimization algorithm [22], [23] is adopted in this paper. Basic backbone of many kinds of Bayesian optimization algorithms follows sequential model-based optimization (SMBO) [22]. In SMBO algorithm, a surrogate model to select optimal hyperparameter candidates is designed considering conditional PDF of observed hyperparameter set. By minimizing or maximizing the output of current surrogate, the optimal hyperparameters in current iteration are determined. After calculating the loss function using such hyperparameters, the loss and corresponding hyperparameters are updated in observed hyperparameter set. Based on this updated hyperparameter set, the surrogate model is newly updated to obtain optimal hyperparameter set in next iteration. All these processes are repeated until the total iteration number satisfies specific threshold. Such SMBO-based Bayesian scheme can be used for the hyperparameter optimization problems of various learning models due to its fast convergence and robustness to model complexity [22]. Especially since both the XGBoost model and Bayesian optimization process are capable of parallel and distributed computing, combination of these two processes has much synergy compared to other learning machines [23]. The overall flow chart of hyperparameter optimization in this paper is described in Fig. 4. The training dataset is generated by the simulation model described in previous sections. Specifically, one training data pair includes 8 by 1 input vector composed of depth and amplitude of cross correlation with 4 different modulation frequencies with MPI, and its corresponding true depth as target value. The loss of optimization in Fig. 4 is same as (35). Such optimization iteration is repeated until $t$ exceeds $\mu_{th}$ which is 30 in this paper. During surrogate model update, the conditional PDF is assumed as Parzen estimator [23]. After the TPE-based hyperparameter optimization is finished, the final optimized hyperparameters of XGBoost are selected as shown in Fig. 4. The detailed training process and validation results are described in the following section.

TABLE I
PARAMETERS FOR AMCW SCANNING LiDAR SIMULATION

| parameter | value | parameter | value |
|---|---|---|---|
| $\Gamma_R$ | 0.1794 V² | $\eta$ | 0.67 |
| $\alpha$ | 1 | $M$ | 50 |
| $f$ | 12.5, 18.75, 25, 31.25 MHz | $F$ | 4.862 |
| $c$ | $3 \cdot 10^8$ m/s | $q$ | $1.60217663 \cdot 10^{-19}$ C |
| $d_{AS}$ | 1.4 ~ 2.4 m | $P_A$ | 0.7854 mm² |
| $m$ | 0.4785 V | $I_{FM}$ | 1 nA/cm² |
| $G$ | 50 kV/A | $T$ | 297 K |
| $\rho_{XYZ}$ | 0 ~ 1 | $E_g$ | 1.1116 eV |
| $d_{AB}$ | 0 ~ 15 cm | $k_B$ | $1.380649 \cdot 10^{-23}$ J/K |
| $t_{transit}$ | 6 ns | $BW$ | 50 MHz |
| $\lambda$ | 852 nm | $S_{TIA}$ | $4.314 \cdot 10^{-24}$ A²/Hz |
| $h$ | $6.62606896 \cdot 10^{-34}$ m² kg/s | $R_{Load}$ | 50 Ω |

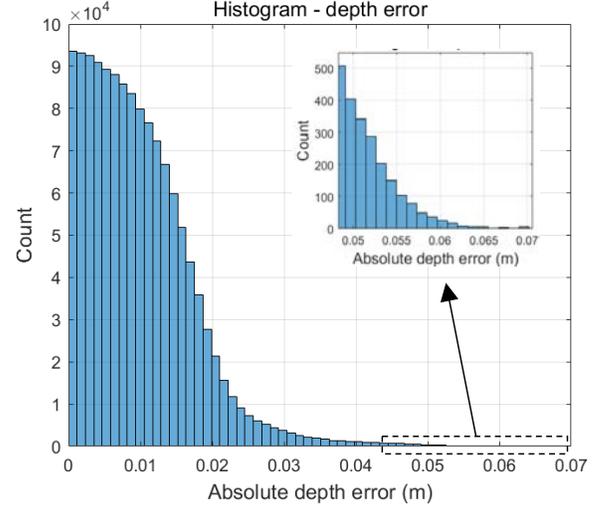

Fig. 5. Absolute depth error distribution due to MPI in simulation dataset.

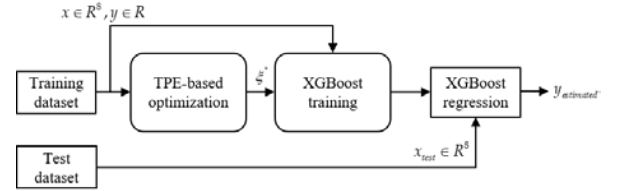

Fig. 6. Block diagram of true depth estimation based on the proposed TPE-optimized XGBoost model.

IV. VALIDATION OF MPI SUPPRESSION METHOD BY SIMULATION AND EXPERIMENT

*A. Validation of MPI suppression based on TPE-optimized XGBoost using MPI simulation data*

To estimate correct depth using TPE-optimized XGBoost model, a precise synthetic training dataset is needed since the acquisition of real MPI data is limited due to the complexity of experimental conditions. To generate the MPI synthetic dataset which reflects the real measurement characteristics of AMCW coaxial-scanning LiDAR, the model parameters in Table I were adopted for the simulation model in this paper. The ranges of the parameters in Table I were widely chosen to sufficiently cover the practical distribution of MPI data. The total number of generated synthetic data is 1,338,670. Each data includes 8 by 1 input vector which includes 4 pairs of depth and amplitude of cross correlation for each modulation frequency (12.5, 18.75, 25, 31.25 MHz), and corresponding scalar output as target (true) depth. Meanwhile, the distribution of absolute depth error due to MPI in simulation dataset is plotted in Fig. 5. The absolute depth error in Fig. 5 is defined as the difference between depth estimation with MPI by simulation model in 31.25 MHz-modulated condition and corresponding true depth. According to the Fig. 5, the mean absolute error (MAE) of depth due to MPI in simulation dataset is about 9.857 mm, which is extremely low compared to conventional ToF sensors [27]. To reduce such MPI error in AMCW scanning LiDAR, the hyperparameter optimization and training regressor is conducted.

Before hyperparameter optimization and training process, the simulation data set is split into 2 parts: 80 % of simulation dataset is selected as training dataset, and 20 % as test dataset. With the training dataset, the TPE-optimized XGBoost is trained. The MPI error correction performance is then validated using the test dataset. All these processes are shown in Fig. 6.

TABLE II
SUMMARY OF DEPTH ERROR CORRECTION
PERFORMANCE INCLUDING MPI

|  | KNN | RF | SVM | DNN | XGBoost |
|---|---|---|---|---|---|
| RMSE-train (mm) | 2.917 | 6.273 | 7.667 | 5.016 | 2.429 |
| RMSE-test (mm) | 3.645 | 6.344 | 7.692 | 4.577 | 2.760 |
| MAE-train (mm) | 1.926 | 4.675 | 5.597 | 3.535 | 1.759 |
| MAE-test (mm) | 2.467 | 4.727 | 5.621 | 3.548 | 1.960 |
| Training time (sec) | 1.272 | 54.93 | 95570 | 2838 | 267 |
| Hardware | CPU | CPU | CPU | GPU | CPU |

To evaluate the MPI suppression performance of TPE-optimized XGBoost more precisely, the comparison study with other widely used learning methods were also examined as follows: K-nearest neighbor (KNN) [40], random forest (RF) [41], support vector machine (SVM) [42], and deep neural network (DNN) with 2 hidden layers and LeakyReLU activation function [43]. As performance indices, root mean square error (RMSE) and mean absolute error (MAE) between target depth and estimated depth were utilized. Meanwhile, to evaluate the model complexity, training time was also compared for each learning-based regression model. The performance comparison for each regression model was conducted in both training and test datasets. All the aforementioned validation results are shown in Table II. The hardware used for the training of KNN, RF, SVM, and XGBoost in Table II is Intel core i9-10900 with 16 GB RAM. Training of KNN, RF, and SVM were processed using python 3.8.8 with scikit-learn library version 1.0.2. For XGBoost training, python 3.8.8 with xgboost library version 1.6.0 was used. For the DNN training, the Quadro RTX 5000 manufactured by NVIDIA was used in same python environment. The library used for DNN training is PyTorch version 1.5.2. Including the XGBoost, other methods except for SVM were also optimized by TPE-Bayesian scheme described in Fig. 4. The optimization tool used in this paper is Optuna [24].

According to Table II, the XGBoost shows the lowest RMSE and MAE in both training and test dataset compared to other learning regressions. Specifically in training dataset, the RMSE and MAE of XGBoost are 2.429 mm and 1.759 mm, respectively. In test dataset, RMSE and MAE of XGBoost are 2.760 mm and 1.960 mm, respectively. Compared to the MAE in Fig. 5 originally 9.857 mm, about 80.12 % reduction of MAE was achieved by the XGBoost-based MPI correction in test dataset. Meanwhile, the RMSE difference between training dataset and test dataset is 0.331 mm, and MAE difference between training dataset and test dataset is 0.201 mm. These relatively small index variations indicate that the training process was conducted without overfitting.

The KNN method also shows low MAE in test dataset about 2.467 mm. This performance is better than that of RF, SVM, and even DNN, which is mainly attributed to the dense distribution of dataset. Especially, the training time of KNN is less than other regression methods due to the simplicity of model. However, the index variation between training and test dataset in MAE is over 0.5 mm which is much larger than that of XGBoost. This means that the KNN is relatively subject to the overfitting compared to other methods. The most inefficient method in Table II is SVM. The training time of SVM is over 24 hours which is extremely long compared to other methods. Such result is mainly due to the extremely many number of memorized support vectors. Moreover, the RMSE and MAE of SVM are larger than those of any other methods. Considering such problems related to time and performance, utilizing SVM is not reasonable for the MPI suppression problem in this paper. The remaining methods, RF and DNN, show moderate MAE and RMSE compared to those of SVM. Specifically, the MAE of RF and DNN in test dataset are 4.727 mm and 3.548 mm, respectively. One attracting point of RF is that the training time is under 60 seconds which is quite less than that of XGBoost. As all weak learners of RF are trained basically in parallel, the total training time can be less than that of XGBoost which depends on sequential-learner training structure. However, the difference of training time between RF and XGBoost is about 213 seconds which is tolerable in general. Such relatively tolerable training time difference is attributed to the parallel training process in each CART and distributing computing of XGBoost [21].

Regarding aforementioned characteristics of each learning regression method, it is easily deduced that utilizing the XGBoost is the most reasonable choice in terms of training time and performance.

*B. MPI suppression of real 3D depth map measured by AMCW coaxial-scanning LiDAR*

To experimentally validate the MPI suppression performance of the proposed XGBoost-based regression model in Fig. 4 and 6, actual object scene including 4 depth maps and 4 amplitude maps were measured by AMCW scanning LiDAR based on parallel-phase demodulation [8]. To acquire obvious MPI error data, multiple corners and sharp points were intentionally chosen as object scenes as shown in Fig. 7(a) and 8(a). The physical quantities of LiDAR are as follows: multiple modulation frequencies of 12.5, 18.75, 25, 31.25 MHz, laser optical power of 20 mW, and the integration time of 16 μsec. For measurement environment, the smooth wooden structure was constructed following the experimental setup of Agresti *et al.* [16] and Buratto *et al.* [27] for the comparison with other previous works. Based on aforementioned experimental conditions, various depth and amplitude maps including MPI error were captured as Fig. 7 and 8. All raw depth map and amplitude map in Fig. 7 and 8 were measured in 31.25 MHz modulation frequency condition.

The depth and amplitude maps of cornered white board in Fig. 7(a) was captured at the distance from 2.05 m to 2.25 m as shown in Fig. 7(b) and (d), respectively. Fig. 7(c) shows the resulting corrected depth map by the proposed method. To acquire depth error maps, the reference map acquired by the geometric information was subtracted by each raw measured depth map and corrected depth map as shown in Fig. 7(e) and (f), respectively. According to Fig. 7(e) and (f), the MPI depth error is mainly distributed around the corner of white board with vertical contour due to the MPI mechanism corresponding to the object's geometry. The maximum absolute depth error is about 2.4 cm in Fig. 7(e). Additionally, the right region of corner much suffers from the MPI error compared to the left region. Such tendency is due to the relatively low intensity of directly reflected light as shown in Fig. 7(d). However, the magnitude of MPI error decreases as the pixel region is getting far from the corner in Fig. 7(e). Such MPI error is drastically suppressed by the proposed TPE-optimized XGBoost as shown in Fig. 7(f), although the noise-like patterns due to random shot noise still exist. Specifically, the MAE of Fig. 7(e) which is 5.0 mm is reduced to 2.6 mm in Fig. 7(f). The original maximum absolute depth error of 2.4 cm is also reduced to 1.1 cm.

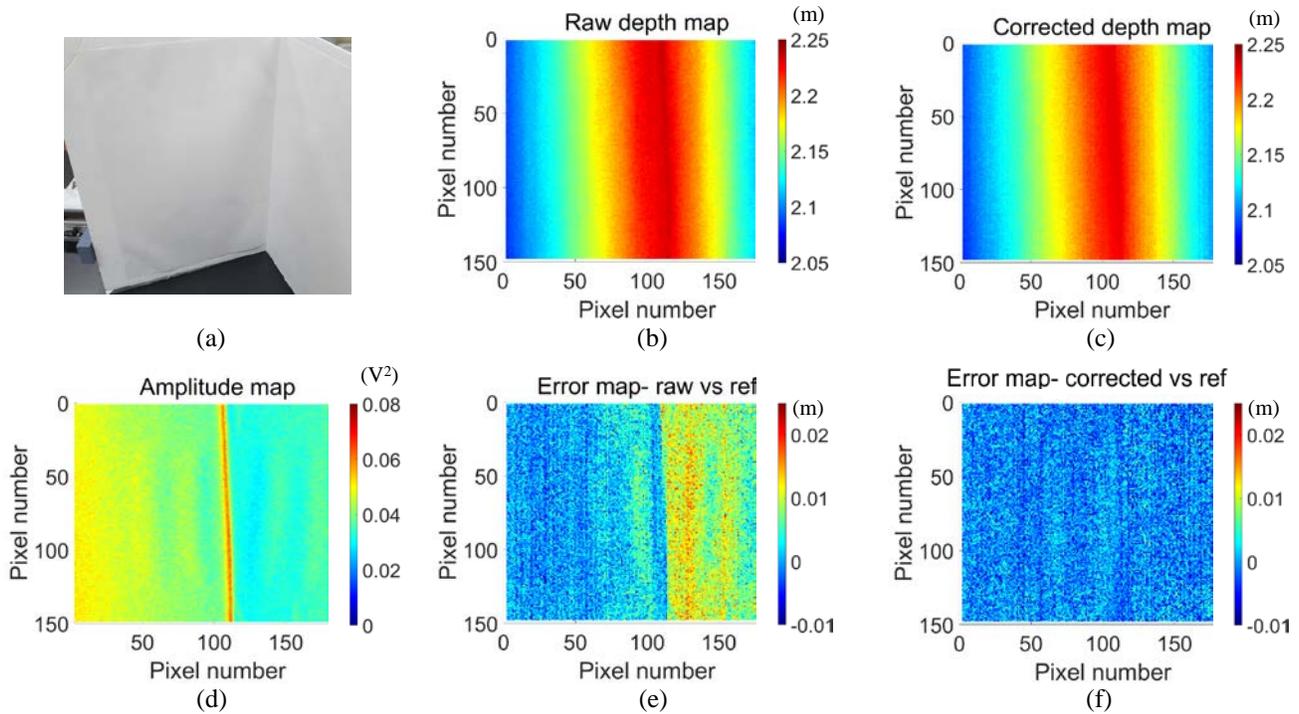

Fig. 7. Depth and amplitude map of cornered paper: (a) object, (b) raw depth map (31.25 MHz), (c) corrected depth map, (d) amplitude map (31.25 MHz) (e) depth error map of raw data, (f) depth error map of corrected data.

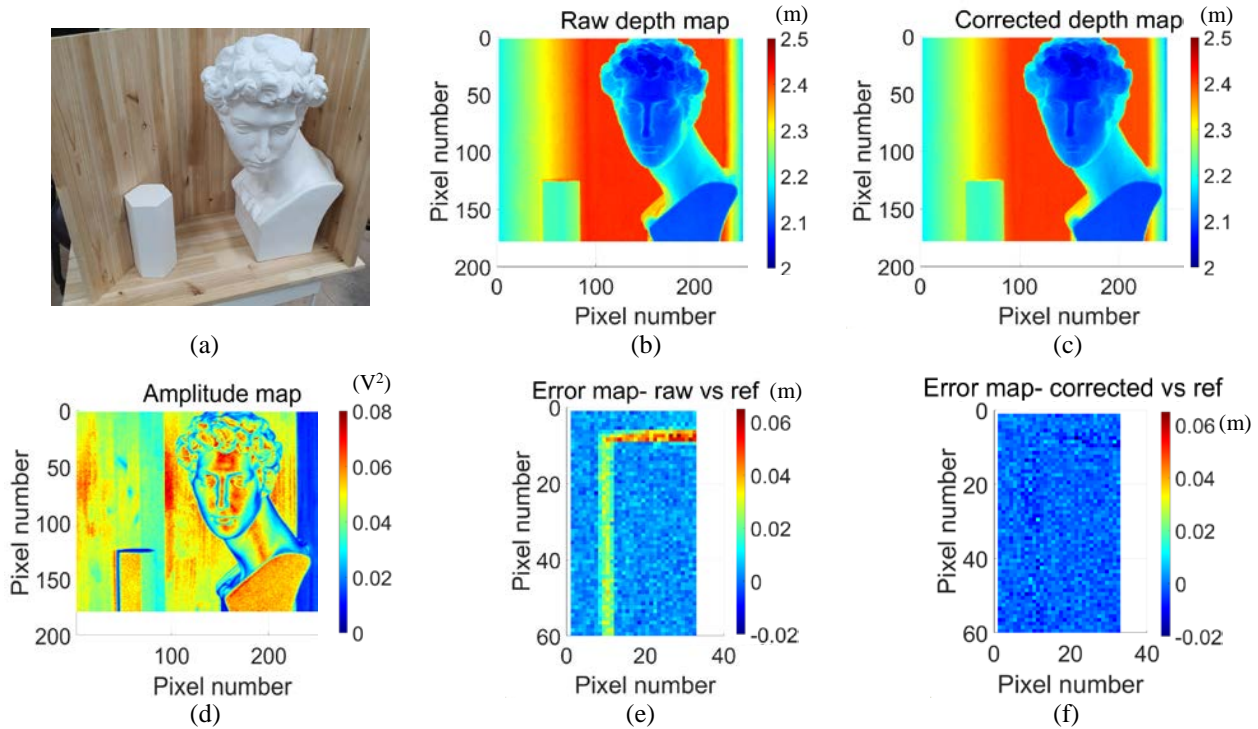

Fig. 8. Depth and amplitude map of multi-objects: (a) object, (b) raw depth map (31.25 MHz), (c) corrected depth map, (d) amplitude map (31.25 MHz) (e) depth error map of raw data, (f) depth error map of corrected data.

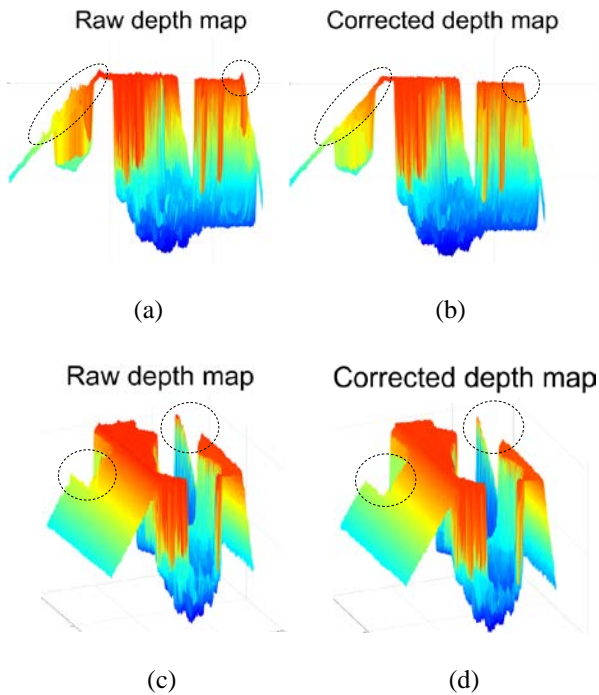

Fig. 9. Depth map of multi-objects: (a) raw data- top view, (b) corrected data- top view, (c) raw data- side view, (d) corrected data- side view.

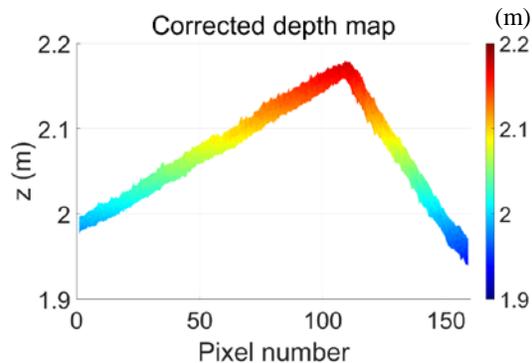

Fig. 10. Depth map of wooden structure- top view.

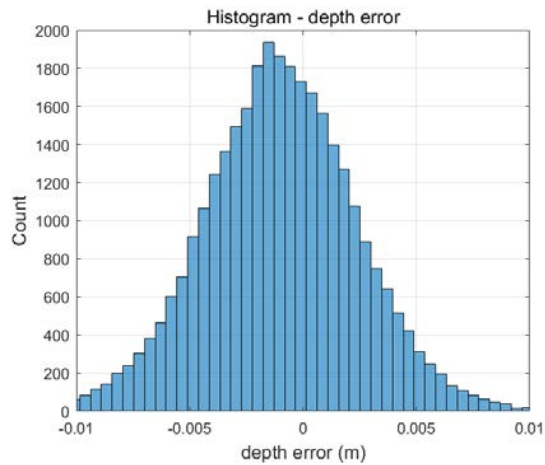

Fig. 11. Histogram of depth error in Fig. 10.

To analyze MPI error distribution in a general complex scene, multi-objects collocated in wooden structure were captured by AMCW scanning LiDAR at distance over 2 m as shown in Fig. 8. In Fig. 8(b), abrupt depth variation due to MPI exists around the wooden corners, left-bottom clavicle of Julien bust, and boundary of hexagonal pillar adjacent to wooden background. Compared to the raw depth map in Fig. 8(b), the corrected depth map in Fig. 8(c) shows much more continuous depth variation in the aforementioned regions, which is close to the actual geometry of the objects. For the quantitative comparison, the depth error maps were also acquired using both raw and corrected depth maps. However, as the Julien bust is geometrically complex, the reference map was generated only in geometrically simple region, as shown in Fig. 8(e) and (f). The maximum absolute depth error and the MAE in Fig. 8(e) are about 6.4 cm and 8.5 mm respectively, which are larger than those in Fig. 7(e). These relatively large MPI errors in Fig. 8(e) are generated since the intensity of directly reflected light in MPI region of Fig. 8(d) (blue-color) is extremely lower than that of Fig. 7(d). Such various SNR of directly reflected light to MPI can be affected by many causes such as the reflectivity of objects, relative position of object and sensor, laser illumination, and the orientation of surface, etc. Namely, it can be deduced that the maximum absolute depth error and MAE in Fig. 8(e) are larger than those in Fig. 7(e) due to the aforementioned various reasons which are not quantitatively presentable though. In Fig. 8(f), the maximum absolute error and MAE are 2.1 cm and 3 mm, respectively. Although there still exist some errors around the boundary of object scene, the majority of depth error was significantly suppressed as shown in Fig. 8(f). Such precise MPI suppression can be also identified in Fig. 9. As shown in Fig. 9, the entire scenes were captured in each top and side view using both raw and corrected depth maps to present the geometrical distortions caused by MPI. The geometrical distortions due to MPI marked by black dotted circle in Fig. 9(a) and (c) were effectively mitigated retaining the actual geometry as shown in Fig. 9(b) and (d).

As a quantitative evaluation of the proposed method, the comparison of the proposed MPI correction method with other previous works for MPI suppression was also accomplished. Unfortunately, due to the different optical structure and modulation frequencies, the public dataset is not compatible with the proposed system and methods in this paper. Alternatively, to compare measurement precision of the proposed method with other previous works, the experimental condition was maintained similar with that of other previous works. Specifically, the wooden structure same as that of Agresti *et al.* [16] and Buratto *et al.* [27] was used for the comparison. As shown in Fig. 10, the corner is almost 90 degree and each wood plate is shown as flat. Specifically, the MAE in Fig. 10 is about 2.8 mm, which is extremely low compared to that of other previous works within cm scale [27]. Additionally, the distribution of depth error is described in Fig. 11. According to Fig. 11, the average bias of depth error is about 1 mm which is negligible. This means that the corrected depth map almost perfectly retains the geometrical information of measured scene even at much longer distance over 2 m compared to other previous works [27].

## V. CONCLUSION

In this paper, the pixel-wise learning approach based on TPE-optimized XGBoost algorithm was proposed along with precise MPI simulation model to correct depth error caused by MPI in AMCW coaxial scanning LiDAR. To optimize the hyperparameter and train the XGBoost, the proposed MPI simulation dataset with practical parameters in Table I was generated and utilized as training and test dataset. According to the test results in Table II, the MAE after MPI correction was about 1.960 mm in test dataset. Such extremely low MAE of depth map was also maintained in real measured object scenes as shown in Fig. 7 to 11. Specifically, the MAE after correction of depth map was lower than 3 mm in various object scenes. To compare the proposed method with other previous works, a similar object was measured and analyzed in Fig. 10 and 11. Consequently, the MAE of the corrected depth map was about 2.8 mm, which is extremely low compared to that of other previous works within cm scale [27]. Such highly precise MPI correction performance of pixel-wise XGBoost regression is mainly attributed to two main characteristics as follows:
1) The inherently low MPI error due to optical characteristics of coaxial scanning LiDAR.
2) Training dataset with four modulation frequencies generated by customized precise simulation model of AMCW coaxial scanning LiDAR.

For the future works, data generation scheme will be additionally improved to further reduce the MPI error in Fig. 8(f). Specifically, domain adaptation will be added to increase the precise training data [16]. Meanwhile, the hyperparameter optimization algorithm can be also modified using other kinds of architectures such as genetic algorithm [44]. Based on these improvements of MPI suppression algorithm, it is anticipated that this work will be utilized in various kinds of scanning LiDAR for the enhancement of 3D depth image quality mitigating effect of MPI.